\documentclass[aps,pre,onecolumn,10pt,longbibliography]{revtex4-2}
\usepackage{geometry}
\usepackage{blindtext}
\usepackage{amsmath,amsthm,amsfonts,amssymb,amscd}
\usepackage{verbatim}
\usepackage{mathrsfs}
\usepackage{natbib}
\usepackage[svgnames,x11names,dvipsnames]{xcolor}
\usepackage{graphicx}
\usepackage{listings}
\usepackage{pgfplots}
\usetikzlibrary{intersections, pgfplots.fillbetween}
\usepackage[colorlinks=true,linkcolor=blue,allcolors=blue,citecolor=blue]{hyperref}
\usepackage{tikz}
\usepackage{physics}
\usepackage{epigraph}  
\usepackage{float}
\usepackage{natbib}
\usepackage{caption}
\usepackage{subcaption}
\usepackage{bm}
\usepackage[capitalise]{cleveref}
\usepackage[export]{adjustbox}
\newcommand{\pd}{\partial}
\newcommand{\ep}{\epsilon}
\newcommand{\wrt}{\mathrm{d}}
\newcommand{\vel}{\textbf{u}}

\newcommand{\ds}{\wrt \text{S}}
\newcommand{\impart}{\text{Im}}
\newcommand{\rar}{\rightarrow}
\newcommand{\zb}{\overline{z}}
\newcommand{\ci}{\mathrm{i}}

\renewcommand{\selectlanguage}[1]{}

\begin{document}
\title{Stokes flow around two unequal cylinders: A complex variable approach}
\author{Luke Neville}
\email{luke.neville@bristol.ac.uk}
\affiliation{School of Mathematics, Fry Building, University of Bristol, BS8 1UG, UK}

\begin{abstract}
We solve the Stokes equations for the flow around two parallel translating and rotating cylinders using tools from complex analysis and conformal mapping. By considering cylinders of arbitrary size and separation, we generalise the solutions known for cylinders of equal size, and a cylinder near a plane wall. We then examine the limit when the cylinders are brought into contact, finding that it affects the separation points in the flow. Namely, they move to, and are hidden in, the point of contact between the cylinders.
\end{abstract}
\maketitle

\section{Introduction}
Multi-body hydrodynamic interactions play an important role in many areas of physics, ranging from colloid and polymer science \cite{doi1988theory,batchelor1972hydrodynamic,batchelor1982sedimentation}, to the design of artificial micro-swimmers \cite{najafi2004simple,golestanian2007designing,lauga2009hydrodynamics}. 
Although the general many-body problem is analytically intractable, simpler systems, such as two spheres or two cylinders can be solved analytically \cite{kim2013microhydrodynamics,happel2012low}. Not only do these solutions give insight into the structure of the full many body problem, they are extremely helpful when checking numerical methods.
 
In this paper we consider the simplest two-dimensional two-body problem, that of two rigidly moving parallel cylinders. Although this problem is simple to state, there are surprisingly few results about it, with solutions only having been found for specific cases. For example, Jeffrey's solution for two counter rotating cylinders \cite{jeffery1922rotation}, and Wakiya's for the dragging apart of two identical cylinders \cite{wakiyabipolar1,wakiya2}. Our purpose here is to show that these solutions exist as limiting cases of a general solution for the flow around two cylinders of \textit{unequal} size. Taking limits, we show that this solution contains all other known two-cylinder solutions, as well as Jeffrey and Onishi's result for a cylinder near a planar wall \cite{jeffrey1981slow}. For completeness we also study two touching cylinders and a cylinder touching a planar wall, for which only the latter has been studied before in the context of matched asymptotic analysis \cite{schubert1967viscous,merlen2011cylinder,terrington2023inner,neville2024controlling}. 

To find this general solution we don't use Jeffery's method of Bipolar coordinates \cite{jeffery1922rotation,wakiyabipolar1}; rather, we take full advantage of the two dimensional setting and use the complex variable formulation of the Stokes equations. These methods began in the field of plane elasticity \cite{muskhelishvili1953some}, and  have repeatedly shown their power when applied to Stokes flows \cite{richardson1968two,hopper1990plane,jeong1992free,leshansky2008surface}. In particular, they were recently used by \citet{crowdy2011treadmilling} and \citet{leshansky2008surface} to re-solve for Jeffrey and Onishi's solution for a cylinder near a wall, and Jeffrey's solution for rotating cylinders respectively. For us, as in these earlier works, the key tool is conformal mapping, as it lets us map all the different two cylinder geometries onto the same annulus.

The rest of the paper is structured as follows. In Section \ref{sec:two cyl} we introduce the problem and solve for the stream function around two rigidly moving unequal cylinders. In Section \ref{sec: touching} we consider the limit when the cylinders are brought into contact, finding that care must be taken as the conformally mapped coordinates become degenerate. We then discuss the effects of this limit further in Section \ref{sec:topology}, before giving some concluding remarks in Section \ref{sec:discuss}.

\section{Two-cylinder problem}
\label{sec:two cyl}

\begin{figure}[h]
\centering
\begin{tikzpicture}[scale=0.6]
\filldraw[color=blue!5,fill=blue!5] (-1,-1) rectangle (12,8); 
\filldraw[color=white,fill=white] (-1,-1) rectangle (10.5,-1.5); 
\filldraw[color=blue, fill=white,line width=0.5mm] (2,3) circle (2);
\filldraw[color=Mahogany, fill=white,line width=0.5mm] (8,3) circle (3);
\draw[-stealth,color=blue, line width=.5mm] (2,5.5) arc (90:140:1.9);
\draw[-stealth,color=blue, line width=.5mm] (8,6.5) arc (90:140:3);
\draw[stealth-stealth,thick] (2,3) -- (8,3);
\draw[stealth-stealth,thick] (-.5,1) -- (-.5,5);
\draw[stealth-stealth,thick] (11.5,0) -- (11.5,6);
\node at (.7,3){$2 R$};
\node at (10,3){$2mR$};
\node at (1.5,6.2){$\Omega_1$};
\node at (6.5,7){$\Omega_2$};
\node at (4.5,2){$2l$};
\end{tikzpicture}
\caption{Diagram of the two cylinder problem. The translational and rotational velocities of each cylinder $(\textbf{U},\Omega)$ are chosen independently. The radius of the left cylinder is $R$, the right is $mR$, and the distance between the two cylinders is $2lR$.}
\label{fig: two cylinders}
\end{figure}
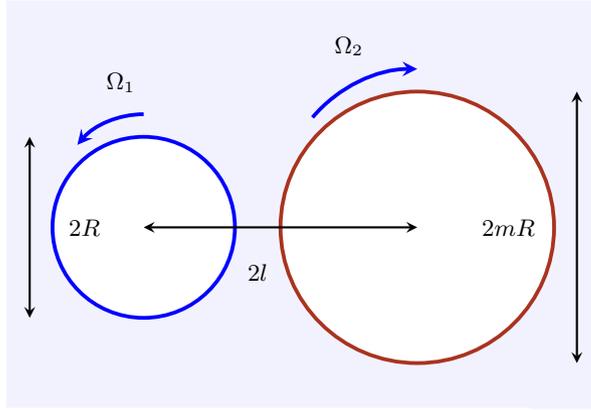

Consider two parallel cylinders of radii $R$ and $2mR$ in an incompressible fluid, separated at a distance $2lR$ as measured from their centres (\cref{fig: two cylinders}). Labelling the left cylinder with one and the right with two, the cylinders are made to rigidly translate at velocity $\textbf{U}_{1,2}$, and rotate about their centre at a speed $\Omega_{1,2}$. The no-slip condition on the cylinders causes the surrounding fluid to be set in motion, and if the Reynolds number is low we can find the fluid velocity with the incompressible Stokes equations
\begin{equation}
    \eta\nabla^2\vel-\bm\nabla p=\textbf{0},\quad \nabla\cdot\vel=0,
\label{eqn:Stokes equations}
\end{equation}
where $\eta$ is the dynamic viscosity, $\vel=(u,v)$ is the velocity, and $p$ is the pressure. As boundary conditions we have
\begin{equation}
    \begin{gathered}
        \vel=\textbf{U}_1 +\Omega_1 \textbf{e}_z\times \textbf{x}\quad  \text{on the left cylinder},\\
        \vel = \textbf{U}_2 +\Omega_2 \textbf{e}_z\times (\textbf{x}-2l\textbf{e}_x)\quad \text{on the right cylinder},\\
        \vel \rar \textbf{0}\ \ \text{at infinity},
    \end{gathered}
    \label{eqn: boundary conditions}
\end{equation}
where the first two conditions come from no-slip, while the last ensures that the fluid becomes quiescent at infinity. Using $\textbf{x}=(x,y)$ for the coordinates in the plane, the left cylinder is centred at the origin, while the right is at $(2lR,0)$. The symbol $\textbf{e}$ denotes unit vectors, with $\textbf{e}_z$ being the unit normal out of the plane. Although the cylinder speeds $(\textbf{U}_i,\Omega_i)$ are currently independent, we later find them to be highly constrained by the condition at infinity.

From now on we work in units where the radius of the left cylinder is one, and the radius of the right is $m$.

\subsection{Solution}

To solve for the flow we use the standard trick for two-dimensional incompressible flows: we introduce a stream function $\psi$ such that  $\textbf{u}=(\pd_y\psi,-\pd_x\psi)$. Substituting the stream function into the Stokes equations (\ref{eqn:Stokes equations}) and taking the curl we arrive at the Biharmonic equation for $\psi$,
\begin{equation}
    \nabla^4\psi=0,
    \label{eqn:stokes biharmonic}
\end{equation}
which must be solved with boundary conditions (\ref{eqn: boundary conditions}). To deal with the complicated geometry (\cref{fig: two cylinders}), classical solutions to (\ref{eqn:stokes biharmonic}) employed bipolar coordinates \cite{jeffery1922rotation,jeffrey1981slow,wakiyabipolar1}. Here we will take an alternative yet equivalent approach using complex variables, thus taking full advantage of our two-dimensional setting \cite{leshansky2008surface,crowdy2011treadmilling}.

Defining a complex coordinate $z=x+\ci y$, the general solution to the biharmonic equation is known to be \cite{jeong1992free}
\begin{equation}
    \psi=\impart[\zb f(z)+g(z)],
\end{equation}
where over-bars denote complex conjugation. The two, so-called Goursat functions $f,g$ are functions of $z$ alone, and are related to the velocity via
\begin{equation}
u-\ci v = 2\ci\frac{\pd\psi}{\pd z}=\bar{z}f'(z)+g'(z)-\overline{f(z)},
\label{eqn:complex vel goursat formula}
\end{equation}
where primes denote the usual complex derivatives \cite{nehari2012conformal}. Although not directly physical, all meaningful quantities may be expressed using the Goursat functions. For example, the function $f$ contains the pressure and vorticity through the combination \cite{crowdy2011treadmilling}
\begin{equation}
p-\ci\eta \omega = 4\eta f'(z).
\end{equation}
In this formulation, solving the Stokes equations is equivalent to finding two functions, $f,g$ that satisfy the boundary conditions (\ref{eqn: boundary conditions}). Following Refs. \cite{crowdy2011treadmilling,leshansky2008surface,jeong1992free,richardson1968two,hopper1990plane}, this will be done by converting the boundary conditions into functional equations that $f,g$ must satisfy and then guessing a solution. This last step is often easier than one would anticipate.

\begin{figure}[t]
\centering
\begin{tikzpicture}[scale=.9]
\filldraw[color=blue,fill=blue!5, line width=0.5mm](7,1) circle (2.5);
\filldraw[color=Mahogany,fill=white, line width=0.5mm](7,1) circle (1);
\draw [to-to,thick](6,1)--(8,1);
\draw [to-to,thick](10,-1.5)--(10,3.5);
\node at (10.5,1){$2$};
\node at (7,1.3){$2S$};
\filldraw[color=black, fill=black,thick] (8.75,1) circle (.1);
\node at (8.75,1.3){$\zeta=a$};
\node at (7,2.75){$\zeta$ space};
\end{tikzpicture}
\caption{Conformal map of the domain outside the two cylinders. The outer boundary with radius $1$ is equivalent to the left cylinder; the inner boundary with radius $S$ maps to the right cylinder, and the thick black dot at $\zeta=a$ maps to infinity}.
\label{fig:conformal map}
\end{figure}
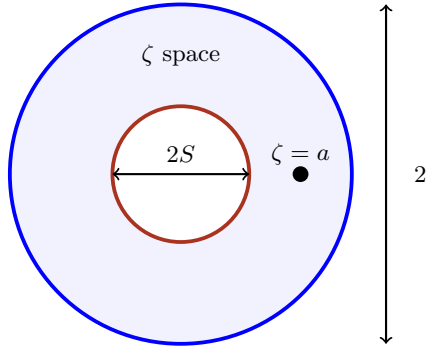

These functional equations could be found in real space, however it is advantageous to conformally map the domain to one with more symmetry. Because the domain is doubly connected (two-holed) it may be mapped to an annulus \cite{nehari2012conformal}. Using $S$ for the inner radius, and setting the outer radius to unity, the required map is given by \cite{koberDictionaryConformalRepresentations1957}
\begin{equation}
\zeta=\frac{1-az}{a-z} \ \Leftrightarrow\ z=\frac{1-a\zeta}{a-\zeta},
\label{eqn:mobius}
\end{equation}
where the parameters $a,S$ related to physical quantities by
\begin{equation}
  (2l-a)(2l-1/a)=m^2,\quad S=\frac{ma}{2l-a},
\label{eqn:parameters}
\end{equation}
and when determining $a$ from the quadratic equation (\ref{eqn:parameters}) we pick the solution that has a magnitude less than one. When the radii of the cylinders are the same we have $m=1$, with $S$ reducing to $a^2$. As shown in \cref{fig:conformal map}, the map (\ref{eqn:mobius}) takes the left cylinder to the outer boundary of the annulus with $|\zeta|=1$, and the right cylinder to the inner one with $|\zeta|=S$. The point at infinity is mapped to the finite value $\zeta=a$, marked as the thick black dot in  \cref{fig:conformal map}. As explained by Crowdy in Ref. \cite{crowdy2020solving}, solving problems by mapping an annulus is equivalent to using bipolar coordinates, and an interested reader could convert all these solutions to bipolar coordinates if needed. 

To find the Goursat functions we convert the complex velocity (\ref{eqn:complex vel goursat formula}) to $\zeta$ coordinates. Employing the chain rule and defining $F(\zeta)=f(z(\zeta)),\ G(\zeta)=\pd_z g(z(\zeta))$ we find
\begin{equation}
    u-\ci v=\overline{z(\zeta)}\frac{F'(\zeta)}{z'(\zeta)}+G(\zeta)-\overline{F(\zeta)}.
\label{eqn:complex vel zeta}
\end{equation}
Using this we convert the boundary conditions on the two cylinders (\ref{eqn: boundary conditions}), to boundary conditions on the inner and outer circles of the annulus. We then replace all $\bar{\zeta}$ dependence in the boundary conditions by noting that for any circle of radius $r$, $\bar{\zeta}=r^2/\zeta$. Using the definition of the Schwarz conjugate $\overline{F}(\zeta)=\overline{F(\overline{\zeta)}}$ this replacement yields  two equations for $F$ and $G$,
\begin{equation}
    \begin{aligned}
        &\frac{\overline{z}(1 / \zeta)}{z^{\prime}(\zeta)} F^{\prime}(\zeta)-\overline{F}(1 / \zeta)+G(\zeta)=\overline{\mathcal{U}}_1-\ci\Omega_1\overline{z}(1/\zeta),\\
            &\frac{\overline{z}\left(S^2 / \zeta\right)}{z^{\prime}(\zeta)} F^{\prime}(\zeta)-\overline{F}\left(S^2 / \zeta\right)+G(\zeta)=\overline{\mathcal{U}}_2-\ci \Omega_2(\overline{z}(S^2/\zeta)-2l),
    \end{aligned}
\label{eqn:collide vel BCs}
\end{equation}
where we have defined complex versions of the rigid body velocities $\mathcal{U}_i=\text{U}_i+\ci\text{V}_i$. 

The two equations (\ref{eqn:collide vel BCs}) have no $\overline{\zeta}$ dependence, and thus are relations between two analytic functions that may be extended from the boundaries of the annulus into the entire domain. Not forgetting the condition at infinity (or $\zeta=a$), solving the Stokes equations has been transformed into finding $F,G$ that satisfy (\ref{eqn:collide vel BCs}). We find it easiest to begin with $F(\zeta)$, taking the difference of the two equations in (\ref{eqn:collide vel BCs}) to  eliminate $G(\zeta)$. Some manipulation of the result yields
\begin{equation}
\begin{aligned}
        &F'(1/\zeta)\frac{(S^2-1)(a\zeta-1)^2}{\zeta(a-\zeta)(a-S^2\zeta)}-\overline{F}(S^2\zeta)+\overline{F}(\zeta)=\overline{\mathcal{U}}_1-\overline{\mathcal{U}}_2
        -\ci\Omega_1\zb(\zeta)+\ci\Omega_2(\zb(S^2\zeta)-2l),
\end{aligned}
\label{eqn:collide stokes functional}
\end{equation}
for which an $F$ must be guessed. Guided by Crowdy's solution to problem of a cylinder moving near a planar wall \cite{crowdy2011treadmilling}, we try the log-polynomial
\begin{equation}
    F(\zeta)=A_1\log\zeta+A_2\zeta+A_3/\zeta,
\end{equation}
where the coefficients are to be determined. Substituting this into (\ref{eqn:collide stokes functional}) and comparing powers of $\zeta$, we arrive at the lengthy expressions for $A_i$ given in Appendix A. The remaining Goursat function $G$, is found by rearranging \textit{either} of the functional equations in (\ref{eqn:collide vel BCs}). Finally, we apply the boundary condition at infinity by taking the limit that $\zeta$ tends to infinity in (\ref{eqn:complex vel zeta}).

\begin{figure}[t]
\centering
\begin{subfigure}[t]{0.03\textwidth}
	\text{(a)}
\end{subfigure}\hfill
\begin{subfigure}[t]{0.45\textwidth}
  \centering
  \begin{adjustbox}{valign=t}
  \includegraphics[width=.8\textwidth]{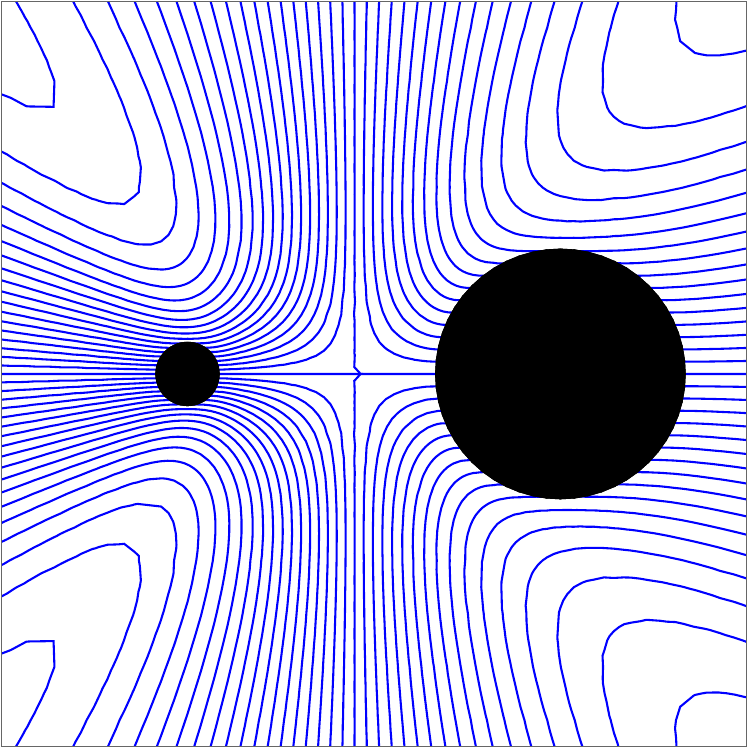}
  \end{adjustbox}
\end{subfigure}\hfill
\begin{subfigure}[t]{0.03\textwidth}
	\text{(b)}
\end{subfigure}\hfill
\begin{subfigure}[t]{.45\textwidth}
  \centering
  \begin{adjustbox}{valign=t}
  \includegraphics[width=.8\textwidth]{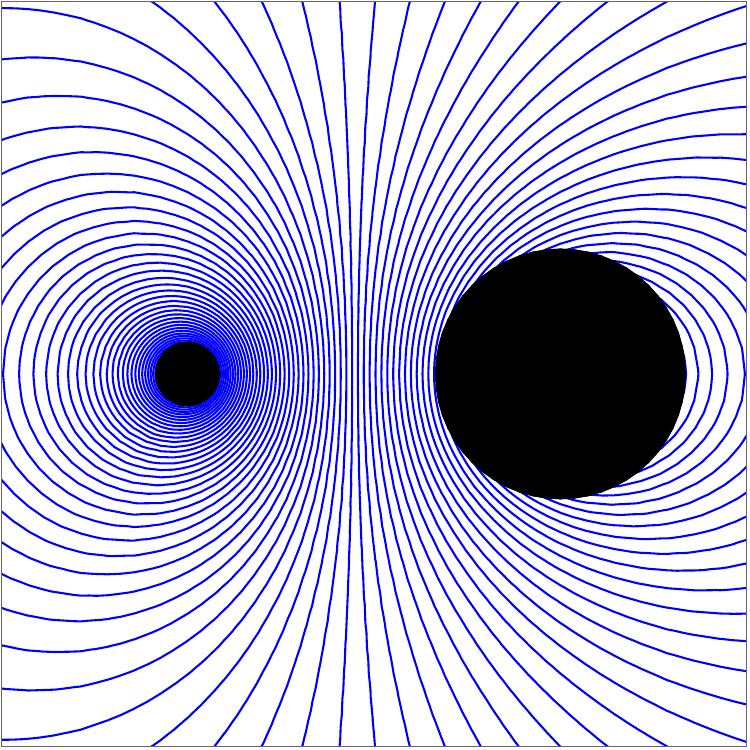}
  \end{adjustbox}
\end{subfigure}
\caption{Streamlines for the Stokes flow around two cylinders where the radius of the right cylinder is four times that of the left. (a) Cylinders moving together along their line of centres (b) Cylinders counter-rotating at speeds such that they both travel vertically at the same speed.}
\label{fig:two flows}
\end{figure}

In this last step we find the constraints on $(\textbf{U}_i,\Omega_i)$ that were mentioned earlier. Namely, we find that $\text{V}_1$ and $\text{V}_2$ depend on the angular velocities, while $\text{U}_2$ is proportional to $\text{U}_1$, with a proportionality factor that decreases as the radius of the right cylinder increases. The general formulae are supplied in Appendix A, but for cylinders of equal radius we have
\begin{equation}
    \begin{gathered}
    \text{U}_1=-\text{U}_2=\text{U}_0/2,\quad
    \text{V}_1+\text{V}_2=\frac{(\Omega_1-\Omega_2)}{2l},\quad
        \text{V}_1-\text{V}_2=\frac{2a(\Omega_1+\Omega_2)\log a}{a^2-1}.\
    \end{gathered}
\label{eqn: swimmer speeds}
\end{equation}
These speeds agree with the solutions found by Wakiya and Jefferey using bipolar coordinates \cite{wakiya2,jeffery1922rotation}, and imply that co-rotating cylinders rotate as a whole about their midpoint, while counter-rotating cylinders self propel. Because the whole system must be torque free, two cylinders rotating clockwise about their centres must also rotate counter-clockwise about a common centre. Streamlines for some examples are shown in \cref{fig:two flows}.

To reduce this solution to that of Jeffery, Onishi, and Crowdy for a cylinder moving near a plane wall \cite{jeffrey1981slow,crowdy2011treadmilling}, we let the radius of the right cylinder tend to infinity. We must ensure that the distance between the left cylinder and wall remains finite in this limit. As such, it is helpful to define a new parameter $h=2l-m$, equal to the distance between the disk and the wall, so that the limit corresponds to taking $m\rar\infty$ with $h$ fixed.

\subsection{Force}
With the solution found we calculate the force and torque on each cylinder. The force, $\textbf{F}$ is given as the integral of the stress tensor $\bm\sigma$ around each cylinder
\begin{equation}
\textbf{F}= \oint_{\text{Left}} \bm\sigma\cdot\textbf{n}\ \ds,
\end{equation}
where $ \ds$ is the unit line element of the cylinder, and $\textbf{n}$ is the unit outward normal. Using the Goursat functions and an identity from Ref. \cite{crowdy2011treadmilling}, this integral is
\begin{equation}
    \text{F}_x+\ci\text{F}_y=2\eta \ci\oint_{|z|=1} \frac{\wrt H}{\ds}\ds,\quad 
    H=f(z)+z \overline{f^{\prime}(z)}+\overline{g^{\prime}(z)},
\label{eqn:complex force H}
\end{equation}
where $(\text{F}_x,\text{F}_y)$ are the force components. Applying the residue theorem yields
\begin{equation}
    \text{F}_x+\ci\text{F}_y=-8\pi\eta A_1,
\label{eqn:force}
\end{equation}
with the force on the right cylinder given by the equal and opposite. To check our calculation against known work we let $m$ tend to infinity to calculate the force on a cylinder near a wall. Doing this we arrive at the expression found by Jeffrey, Onishi, and Crowdy for a cylinder moving near a solid wall \cite{jeffrey1981slow,crowdy2011treadmilling}, confirming our calculation.

\subsection{Torque}

The torque is found similarly, and using (\ref{eqn:complex force H}), the torque on the left cylinder, about its origin, is given by
\begin{equation}
    \text{T} =\oint_{\text{Left}}\ \textbf{x}\times(\bm{\sigma}\cdot\textbf{n})\ \ds=2\eta \text{Re}\left[\oint_{|z|=1}zg''(z){\wrt z}\right].
\end{equation}
Substituting in the solution and applying the residue theorem we find
\begin{equation}
    \text{T} =
\frac{4 \pi \eta\left(S^2(\Omega_1-\Omega_2)+a^4 S^2(\Omega_1-\Omega_2)-a^2\left(\Omega_1+S^4 \Omega_1-2 S^2 \Omega_2\right)\right)}{\left(-1+a^2\right)\left(a^2-S^2\right)\left(-1+S^2\right)},
\label{eqn:torque}
\end{equation}
with the torque on the right equal and opposite. These explicit calculations show that the whole system is force and torque free, as required to avoid the Stokes paradox associated with an isolated cylinder \cite{jeffery1922rotation}. As we show now, this can be seen directly from the far-field behaviour of the Goursat functions.

\subsection{Far-Field behaviour}

Far away from the cylinders we expect the flow to be well approximated by several point singularities whose strength depend on the cylinders' speed and size. These may be found by expanding the Goursat functions for large $z$ and collecting leading order singularities. This expansion yields
\begin{equation}
    \begin{gathered}
        f(z)\sim \frac{\sigma_1}{(z-l)},\quad \sigma_1 = \frac{(a^2-1)(aA_1+a^2 A_2-A_3)}{a^2}\\
        g'(z)\sim \frac{l\sigma_1}{(z-l)^2}+\frac{\sigma_2}{(z-l)^2},\quad
        \sigma_2 = \frac{(1-a^4)\overline{A}_1}{2a^2}+\frac{(1-a^2)\bar{A}_2}{a^3}+(a^3-a)\bar{A}_3+\ci l\Omega_1,
    \end{gathered}
\end{equation}
corresponding to a stresslet of strength $\sigma_1$, and an irrotational dipole of strength $\sigma_2$ centered at the midpoint between each cylinder \cite{pozrikidis1992boundary}. Both these singularities are force and torque free, confirming the direct calculations of the previous statement.

\section{Touching cylinder limit}
\label{sec: touching}
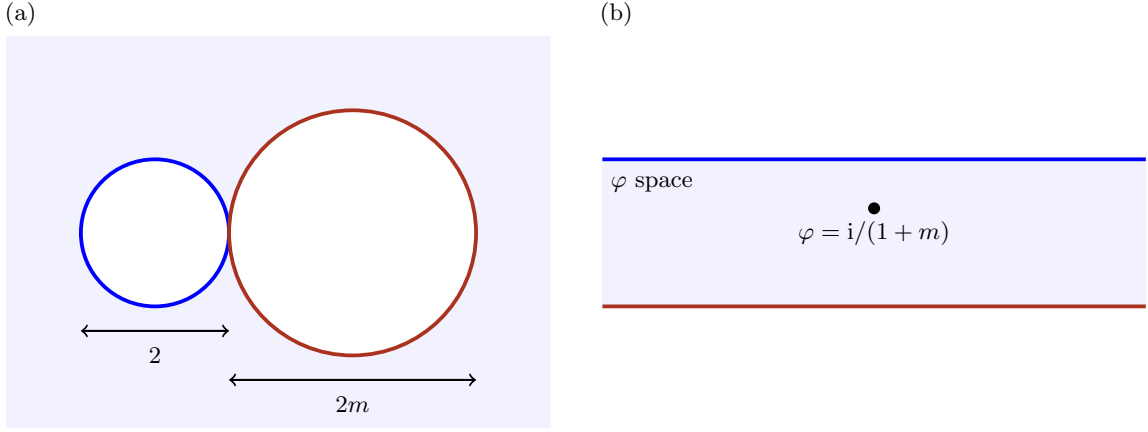
\begin{figure}
\centering
\begin{subfigure}[t]{0.49\textwidth}
	\caption{ }
    \begin{tikzpicture}[scale=0.65]
\filldraw[color=blue!5,fill=blue!5] (-1,0) rectangle (10,8); 
\filldraw[color=white,fill=white] (-1,0) rectangle (10.5,-1.5); 
\filldraw[color=blue, fill=white,line width=0.5mm] (2,4) circle (1.5);
\filldraw[color=Mahogany, fill=white,line width=0.5mm] (6,4) circle (2.5);
\draw [to-to,thick](.5,2)--(3.5,2);
\node at (2,1.5){$2$};
\node at (6,.5){$2m$};
\draw [to-to,thick](3.5,1)--(8.5,1);
\end{tikzpicture}
\end{subfigure}\hfill
\begin{subfigure}[t]{0.49\textwidth}
\caption{ }
\begin{tikzpicture}[scale=.65]
\filldraw[color=white,fill=white] (-1,3) rectangle (10.5,5.5); 
\filldraw[color=blue!5,fill=blue!5] (-1,0) rectangle (10,3);
\filldraw[color=white,fill=white] (-1,0) rectangle (10.5,-1.5); 
\draw [-,color=Mahogany, line width=0.5mm](-1,0)--(10,0);
\draw [-,color=blue, line width=0.5mm](-1,3)--(10,3);
\filldraw[color=black, fill=black,thick] (4.5,2) circle (.1);
\node at (0,2.5){$\varphi\ \text{space}$};
\node at (4.5,1.5){$\varphi=\ci/(1+m)$};
\end{tikzpicture}
\end{subfigure}
\caption{(a) Two touching cylinders. The left cylinder has radius one, while the right has radius $m$. (b) Conformally mapped domain. The thick black dot corresponds to the point at infinity in real space, the top surface to the left cylinder, and the bottom surface to the right cylinder. Regions with the same colour are mapped to each other.}
\label{fig:touching mapping}
\end{figure}

The solution we have just found holds for all cylinder separations and ought to work when we the cylinders are brought into contact. This limit however, has difficulties, one physical, one mathematical. Physically, the force between the cylinders (\ref{eqn:force}) diverges as they are brought together  due to the strong squeezing of fluid between the two \cite{jeffrey1981slow}. Mathematically, we see that the topology of the fluid domain is different in the limit, being doubly connected when the cylinders are apart, but singly connected with they are together \cite{nehari2012conformal}. 

The above facts indicate that we have to be careful in taking this limit. Indeed, a naive application of the limit, which corresponds to setting $a=1$ fails, as the conformal map (\ref{eqn:mobius}) degenerates. Going about this more carefully, we will first bring the cylinders infinitesimally close together, setting $2l = (1+m)(1+\ep^2/2m)$, with $\ep$ a small parameter. Substituting this into the mapping parameters (\ref{eqn:parameters}) gives
\begin{equation}
a = 1-\ep +O(\ep^2), \quad S=1-\ep(1+1/m)+O(\ep^2),
\end{equation}
which correspond to an annulus of thickness $\ep(1+1/m)$. In such a thin annulus $\zeta$ is no longer the appropriate coordinate as it varies only by an $O(\ep)$ amount when we move between the inner and outer circles. A better coordinate is given by $\varphi$, related to $\zeta$ through $\zeta=S+\ep\varphi/m$; we see that $\varphi$ measures radial distance from the inner circle of the annulus. Substituting these into the conformal map (\ref{eqn:mobius}) and taking the limit that $\ep$ tends to zero, we find a new map
\begin{equation}
z = 1+\frac{2m}{1-\varphi},\quad \varphi = 1 +\frac{2m}{1-z},
\end{equation}
which takes the region outside the cylinders to a vertical channel. Although useful, this map fails in the limit of $m\rar\infty$, corresponding to a cylinder touching a flat wall. For this reason we will scale the $\varphi$ coordinate by a factor of $\ci/(1+m)$, as the factor of $(1+m)$ in the denominator is able to balance the factor of $m$ in the numerator, while the factor of $\ci$ is added to rotate the channel. We also find it helpful to shift the $z$ coordinate so that the left cylinder has an origin at $z=-1$. With these changes made, we have \cite{koberDictionaryConformalRepresentations1957}:
\begin{equation}
    \varphi = \frac{\ci}{1+m}\left(1-\frac{2m}{z}\right),
    \label{eqn:touching map}
\end{equation}
which, as shown in \cref{fig:touching mapping}, takes the domain outside the cylinders to the channel:  $\mathfrak{Re}[\varphi]\in (-\infty,\infty)$, $\mathfrak{Im}[\varphi]\in[0,1]$. The boundary of the right cylinder is mapped to the lower surface with $\mathfrak{Im}[\varphi]=0$, and the boundary of the left to the upper surface with $\mathfrak{Im}[\varphi]=1$. The point at infinity goes to $\ci/(1+m)$, and the point of contact between the two maps to infinity. When the right cylinder becomes infinitely big we recover, up to a scale factor, the conformal map used by Schubert to study a cylinder touching a planar wall \cite{schubert1967viscous}. 

We appreciate that the variable changes we made in taking this limit may not be obvious. Moreover, the change in mapped geometry from an annulus to a flat channel may have killed any geometric intuition from the last section. In \cref{fig:map limit} we attempt to explain these steps through a series of pictures.
\begin{figure}
\centering
\begin{subfigure}[t]{0.28\linewidth}
	\caption{ }
    \begin{tikzpicture}[scale=.9]
\filldraw[color=white,fill=white, line width=0.5mm](-3,3) rectangle (3,3.5);
\filldraw[color=blue,fill=blue!5, line width=0.5mm](0,0) circle (3);
\filldraw[color=Mahogany,fill=white, line width=0.5mm](0,0) circle (2.5);
\filldraw[color=black, fill=black,thick] (2.75,0) circle (.1);
\node at (1.2,0){$\zeta= a\approx 1-\ep$};
\node at (0,-1.6){$S\approx 1-\ep(1+m)/m$};
\draw[<->,color=black,thick] (0,-2.5)--(0,-3);
\end{tikzpicture}
\end{subfigure}\hfill
\begin{subfigure}[t]{0.28\textwidth}
	\caption{ }
    \begin{tikzpicture}[scale=0.75]
\draw[-,color=blue, line width=1mm,name path=A] (5.19,-3) arc (-30:30:6);
\draw[-,color=Mahogany, line width=1mm,name path=B] (6.93,-4) arc (-30:30:8);
\tikzfillbetween[of=A and B]{blue!5};
\filldraw[color=black, fill=black,thick] (7.5,0) circle (.12);
\node at (5.5,4){$\ep\ll 1$};
\end{tikzpicture}
\end{subfigure}\hfill
\begin{subfigure}[t]{0.28\textwidth}
	\caption{ }
    \begin{tikzpicture}[scale=0.9]
\filldraw[color=blue!5,fill=blue!5] (-1,-3) rectangle (1,3);
\draw [-,color=Mahogany, line width=0.5mm](-1,-3)--(-1,3);
\draw [-,color=blue, line width=0.5mm](1,-3)--(1,3);
\filldraw[color=black, fill=black,thick] (0.5,0) circle (.1);
\node at (0,3.5){$\ep=0$};
\end{tikzpicture}
\end{subfigure}\hfill

\caption{Series of changes that result in the map (\ref{eqn:touching map}). In all figures the blue and red lines correspond to the same regions, with the thick black dot being the point at infinity. (a) A very thin annulus. The eventual contact point between the cylinders is at $\zeta=-1$, on the opposite side from the point at infinity (b) Zooming in on the thin annulus near the point at infinity we see that it looks flatter and flatter, with the contact point becoming infinitely far away. (c) Taking the limit that $\ep\rar 0$ we arrive at a flat vertical channel. Multiplying this map by $\ci/(1+m)$ results in the conformal map shown in \cref{fig:touching mapping}. The contact point between the two cylinders is at infinity, and may be thought of as a point in an imaginary annulus connecting the two sides of the channel. }
\label{fig:map limit}
\end{figure}

The solutions for the touching cylinders may be found by performing the same variable changes on the Goursat functions, however this procedure does not give intuition for their functional form. For this reason we shall re-solve for the flow around two touching cylinders using the same methods as section \ref{sec:two cyl}. Proceeding as before, we first write the complex velocity in terms of $F(\varphi(z))$ and $G(\varphi(z))$ 
\begin{equation}
    u-\ci v = \frac{(-\ci+\varphi+m \varphi)^2}{(1+m)(\ci+m\varphi+\overline{\varphi})}F'(\varphi)+G(\varphi)-\overline{F(\varphi)}.
\end{equation}
We then transform the two cylinder boundary conditions to boundary conditions in the mapped channel. The left cylinder corresponds to the bottom line in \cref{fig:touching mapping} and so we have $\overline{\varphi}=\varphi$, while the right cylinder is the top line with $\overline{\varphi}=\varphi-2\ci$. The boundary conditions are thus written as
\begin{equation}
    \begin{gathered}
        \begin{aligned}
        &\frac{(-\ci+\varphi+m\varphi)^2}{(1+m)(-\ci+\varphi+m(-2\ci+\varphi))}F'(\varphi)+G(\varphi)-\overline{F}(\varphi-2\ci)=\overline{\mathcal{U}}_1-\ci\Omega_1(\overline{z}(\varphi-2\ci)+1).
    \end{aligned}\\
    \begin{aligned}
        &\frac{(-\ci+\varphi+m \varphi)^2}{(1+m)(\ci+m \varphi+\varphi)}F'(\varphi)+G(\varphi)-\overline{F}(\varphi)=\overline{\mathcal{U}}_2-\ci\Omega_2(\overline{z}(\varphi)-m),
    \end{aligned}
    \end{gathered}
    \label{eqn: touching BC functional}
\end{equation}
Subtracting one relation from the other we arrive at a functional equation for $F(\varphi)$
\begin{equation}
\begin{aligned}
        &\frac{2 \ci(-\ci+\varphi+m \varphi)^2}{(\ci+m\varphi+\varphi)(-\ci+\varphi+m(-2\ci+\varphi))}F'(\varphi)+\overline{F}(\varphi)-\overline{F}(\varphi-2\ci)=\overline{\mathcal{U}}_2-\overline{\mathcal{U}}_1\\
        &-\ci\Omega_2(\overline{z}(\varphi)-m)+\ci\Omega_1(\overline{z}(\varphi-2\ci)+1),
\end{aligned}
\end{equation}
for which a solution must be guessed. The flow must be finite everywhere except the point of contact at $\varphi=\infty$. This means that we cannot have any logarithms or negative powers of $\varphi$ in the Goursat functions as these blow up in the fluid domain. $F$ must then have a Taylor series, and substituting the series we find that only three terms are needed
\begin{equation}
    F(\varphi)=B_1\varphi+B_2\varphi^2+B_3\varphi^3,
\end{equation}
with the coefficients given in Appendix B. As before, the other Goursat function is found using either of the functional equations in (\ref{eqn: touching BC functional}).

\begin{figure}[t]
\centering
\begin{subfigure}{0.45\textwidth}
\caption{ }
  \includegraphics[width=.8\textwidth]{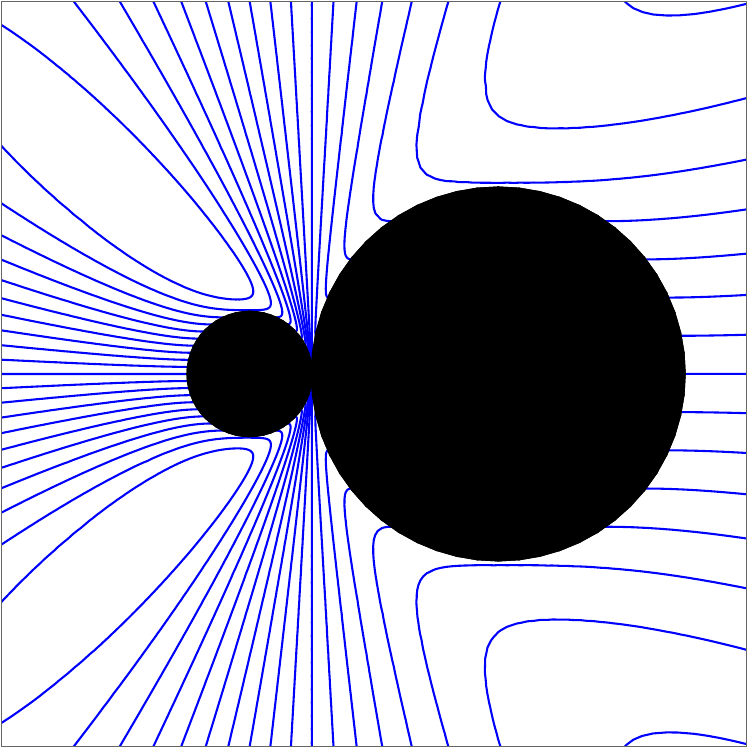}
\end{subfigure}
\hfill
\begin{subfigure}{0.45\textwidth}
\caption{ }
  \includegraphics[width=.8\textwidth]{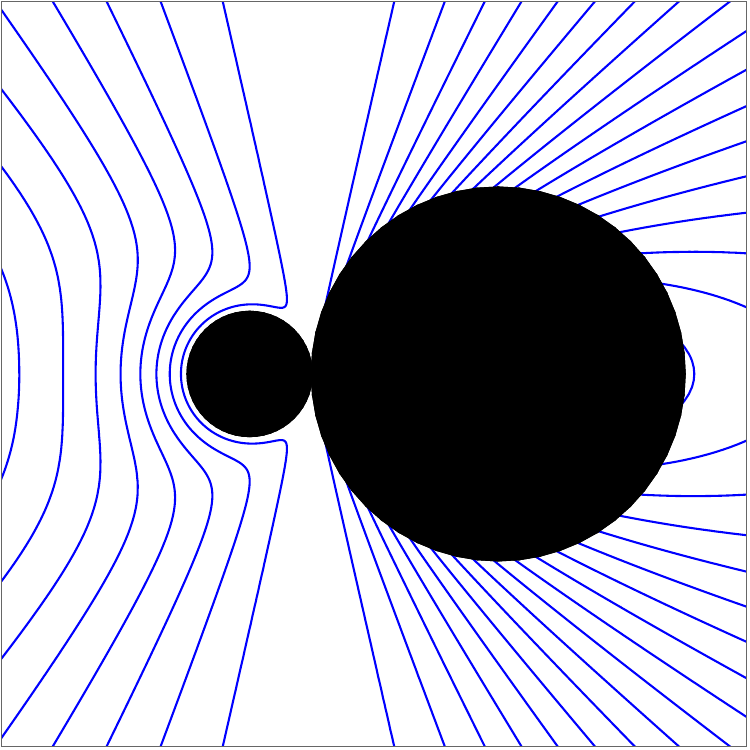}
\end{subfigure}

\begin{subfigure}{0.45\textwidth}
\caption{ }
  \includegraphics[width=.8\textwidth]{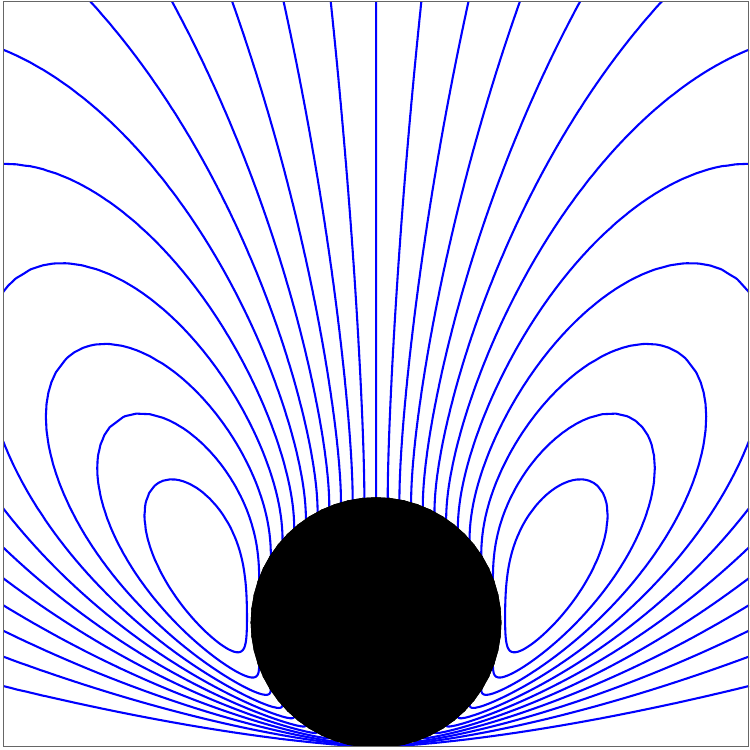}
\end{subfigure}
\hfill
\begin{subfigure}{0.45\textwidth}
\caption{ }
  \includegraphics[width=.8\textwidth]{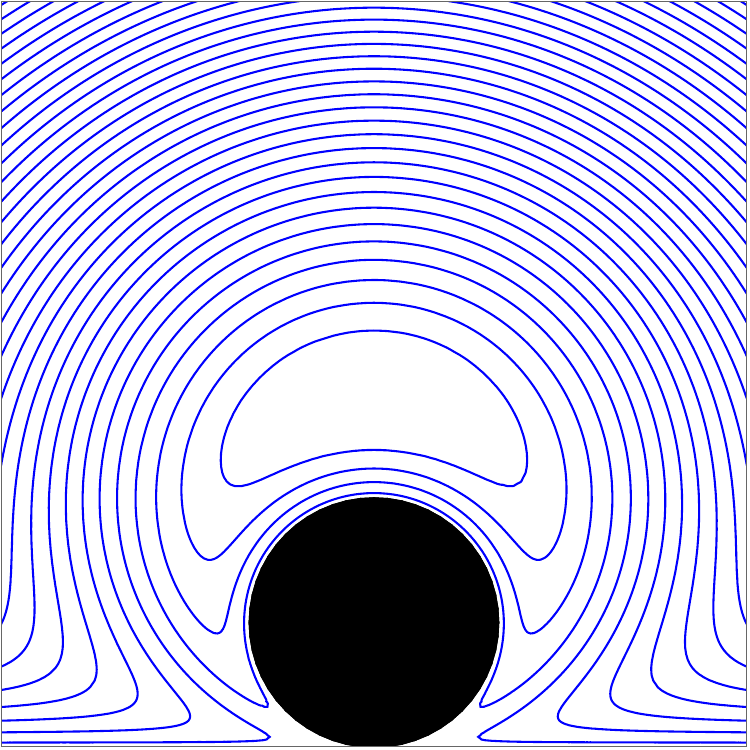}
\end{subfigure}
\caption{Streamlines of the Stokes flows. (a,b) Two cylinders where the right radius is three times the left; (c,d) cylinder touching a wall. (a) Two cylinders moving towards each other along their line of centres. (b) Left cylinder rotates with $\Omega_1$=1 while right rotates and translates at speeds given by (\ref{eqn:touching speeds}). (c) Cylinder moving downwards onto a stationary wall. (d) Cylinder rotating anti-clockwise on a wall that is translating uniformly right. This solution is obtained by boosting into the rest frame of the cylinder.}
\label{fig:touching flow}
\end{figure}

As for non-touching cylinders we find the speeds to be highly constrained. Choosing to fix $\text{U}_1,\text{V}_1,\Omega_1$, we have
\begin{equation}
\begin{gathered}
\text{U}_2=-\frac{\text{U}_1(1+3m)}{(3+m)m^2},\\
\text{V}_2=\frac{\Omega_1(1+m)-\text{V}_1(1+2m)}{m^2},\\
\Omega_2 = \frac{2\text{V}_1(1+m)-(2+m)\Omega_1}{m^2},
\end{gathered}
\label{eqn:touching speeds}
\end{equation}
which all go to zero as the radius of the right cylinder tends to infinity, i.e. when it becomes a stationary plane wall. We have checked the solutions are equivalent to those found by \citet*{schubert1967viscous,merlen2011cylinder,neville2024controlling} in this limit. Some example flows are shown in \cref{fig:touching flow}.

\section{Topological defects in the velocity field}
\label{sec:topology}
\begin{figure}[b]
\centering
\begin{subfigure}{0.45\textwidth}
\caption{ }
  \includegraphics[width=.8\textwidth]{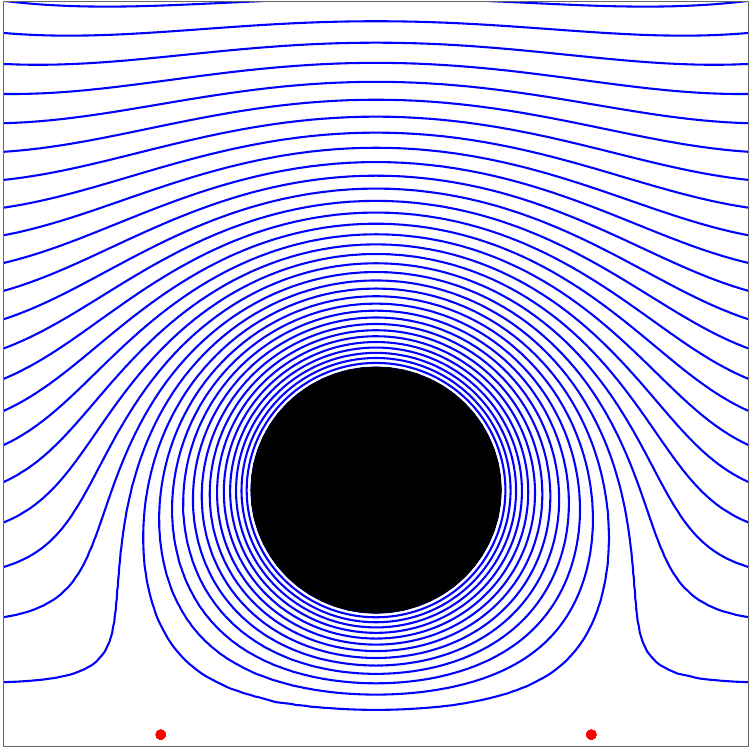}
\end{subfigure}
\hfill
\begin{subfigure}{0.45\textwidth}
\caption{ }
  \includegraphics[width=.8\textwidth]{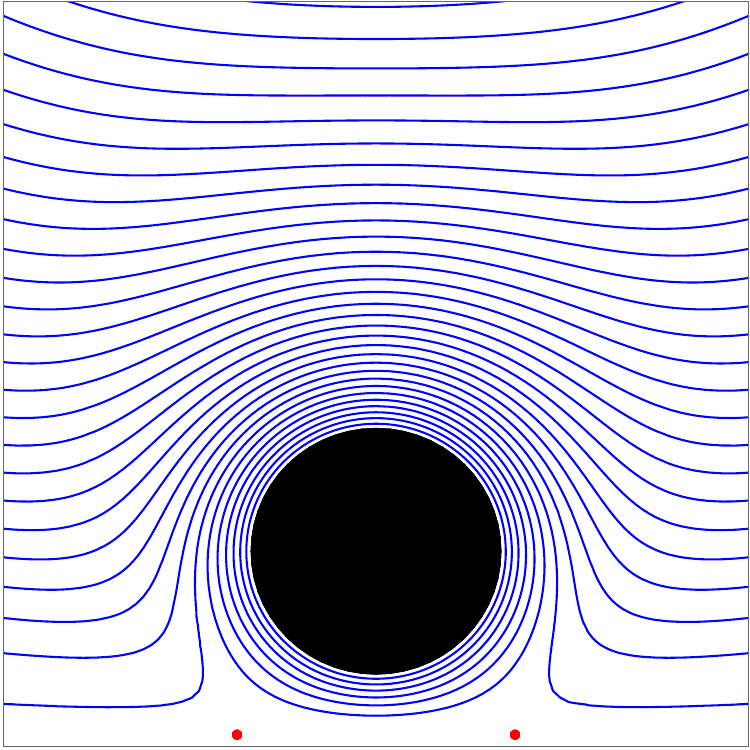}
\end{subfigure}

\begin{subfigure}{0.45\textwidth}
\caption{ }
  \includegraphics[width=.8\textwidth]{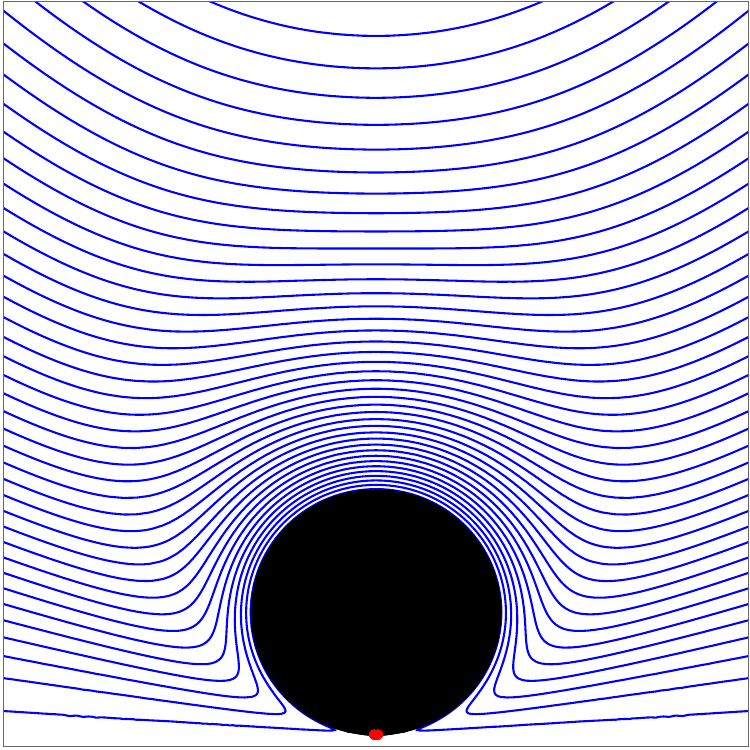}
\end{subfigure}
\hfill
\begin{subfigure}{0.45\textwidth}
\caption{ }
  \includegraphics[width=.8\textwidth]{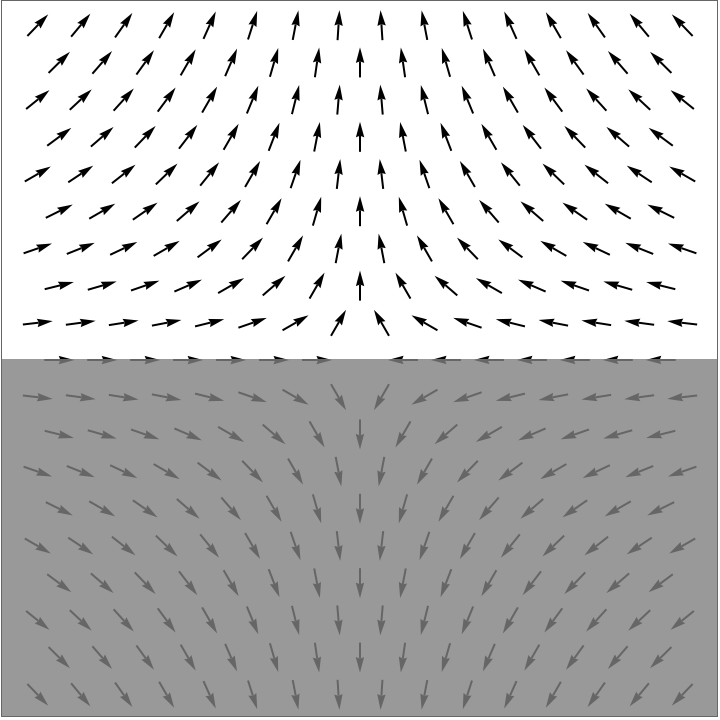}
\end{subfigure}
\caption{(a-c) Streamlines of the Stokes flows around a rotating cylinder with circulation $2\pi$. The red dots mark the separation points. (a-b) As the distance between the wall and cylinder is decreased, so does the distance between the two points. (c) When the cylinders touches the wall the separation points have merged and sit at the point of contact. (d) Schematic of the velocity field around the separation point on the wall. The $-1$ defect in the velocity field is cut in half by the wall (grey region), leading to an effective $-1/2$ defect.}
\label{fig:approaching singularities}
\end{figure}

As mentioned in section \ref{sec: touching}, the problem of two touching cylinders is topologically distinct from the non-touching problem. Here we study the effect of this for systems with non-zero circulation. The circulation $\Gamma$ in some region $\mathcal{D}$ is given as the line integral
\begin{equation}
\Gamma=\oint_{\partial\mathcal{D}}\vel\cdot\textbf{n}\ \ds.
\label{eqn:circulation}
\end{equation}
If the cylinders have non-zero angular velocity then they have non-zero circulation $\Gamma_i =2\pi\Omega_i$. However the boundary condition at infinity (\ref{eqn: boundary conditions}) implies that the whole system is circulation free. For this to be consistent there must be regions of opposite circulation. However, looking at the flow for a rolling cylinder in contact with a wall (\cref{fig:approaching singularities}(d)), there appears to be a contradiction. The cylinder has non-zero circulation, but there appear to be no compensating regions of opposite circulation. They are, in fact, hidden at the point of contact.

To see this we first plot the non-touching case in \cref{fig:approaching singularities}(a-c). We see that the regions of opposite circulation are the the separation points in the flow which sit directly on the wall. Their horizontal positions may be readily identified as
\begin{equation}
x_c = \pm  \frac{1-a^2}{2a},
\label{eqn:singularity coordinates}
\end{equation}
where $a$ is given by (\ref{eqn:parameters}). We plot these as red dots in \cref{fig:approaching singularities}, showing how they move under the cylinder as it approaches the wall. The separation points merge once the cylinder touches the wall and sit directly at the contact point. Therefore when calculating the net circulation the point of contact must be included to avoid the earlier contradiction. 

At this point one may have noticed that the separation points are also topological defects in the velocity field with charge $-1/2$ \cite{degennesprost}. However the velocity field is  a vector field and so defects ought to only come with integer charge \cite{mermin1979topological}. This apparent inconsistency is resolved by thinking of the $-1/2$ defect at the wall is a $-1$ defect cut in half along its symmetry line by the wall (\cref{fig:approaching singularities}(d)).

\begin{figure}[h]
\centering
\begin{subfigure}{0.45\textwidth}
\caption{ }
  \includegraphics[width=.8\textwidth]{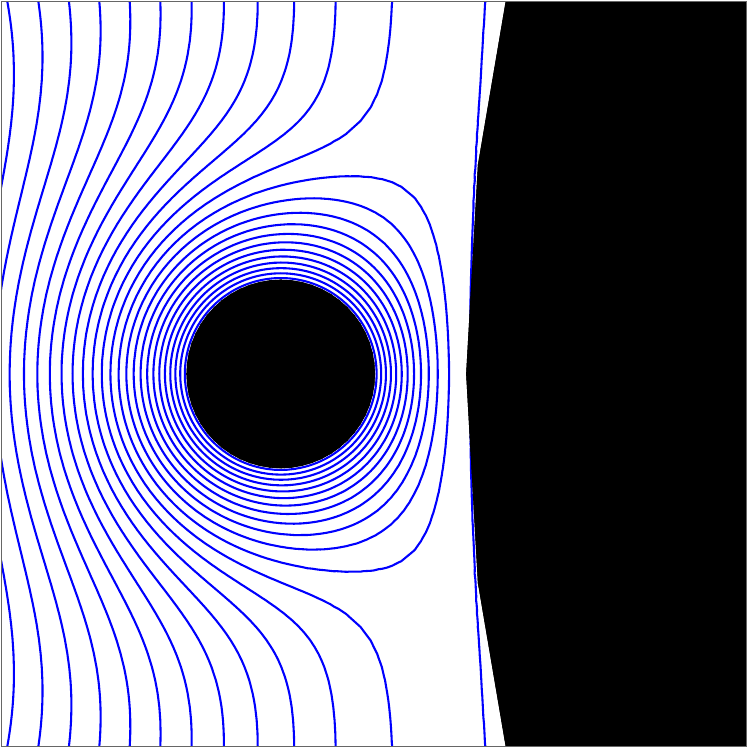}
\end{subfigure}
\hfill
\begin{subfigure}{0.45\textwidth}
\caption{ }
  \includegraphics[width=.8\textwidth]{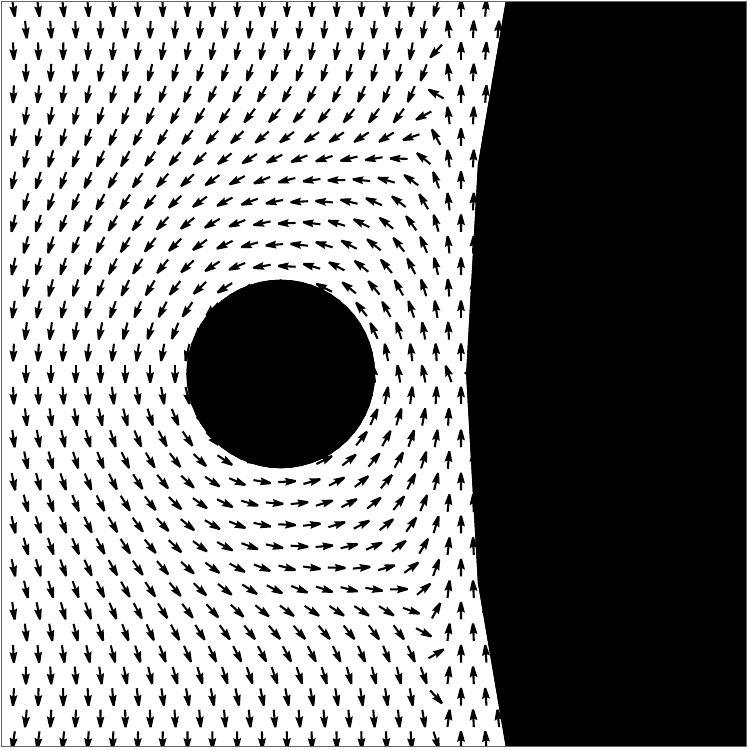}
\end{subfigure}
\caption{Velocity field around two cylinders where the radius of the right is twenty times that of the left. The angular velocity of the left cylinder is set to one while that of the right is given by the formula in Appendix A. (a) Streamlines of the velocity field. (b) Vector field which better shows the vortex stretching.}
\label{fig: Rolling Larger}
\end{figure}

The situation is quite different when the radius of the right cylinder is large but finite. There are no topological defects in the velocity field, and the separation points are replaced by stretched vortices (\cref{fig: Rolling Larger}). The vortices then shrink and move towards the right cylinder as its radius increases, eventually touching it and becoming the separation points mentioned above.

\section{Discussion}
\label{sec:discuss}

In this work we have used complex variable methods to determine the Stokes flow around two rigidly translating and rotating cylinders, generalising work by many authors \cite{jeffery1922rotation,wakiyabipolar1,wakiya2,jeffery1922rotation,jeffrey1981slow,merlen2011cylinder}. While the cylinder speeds were originally independent, while solving the problem we found them to be highly constrained, with the cylinders forced to move as a coordinated pair \cite{wakiyabipolar1}. Calculating the force and torque on each cylinder, we showed that they were equal and opposite on both, with none of the usual paradoxes associated with two-dimensional Stokes flow \cite{wakiyabipolar1,wakiya2,jeffery1922rotation,crowdy2011treadmilling}.

We then brought the cylinders together till they were in contact, finding that the limit required some care, with coordinated having to be scaled together as the limit was taken. Although we were able to take this limit and recover the solutions of Refs. \cite{merlen2011cylinder,schubert1967viscous, terrington2023inner}, we found it beneficial to re-solve the problem using an alternate conformal map based on a map to a channel. Investigating this limit further in section \ref{sec:topology} we found that it affected the separation points in the flow, causing them to move under the cylinder and merge at the point where it touches the wall. We argued that these separation points could be thought of as topological defects in the velocity field. Interestingly we recently found a similar effect in liquid crystal systems \cite{neville2024controlling}, with the topological defects in the nematic texture also moving to the point of contact under the disk. 

Although our analysis focused on two-dimensional flow corresponding to infinitely long cylinders, we would be interesting to tackle the more realistic problem of a finite cylinder. This could be attacked with slender body theory \citep{de1973low,batchelor1970slender}, and it would be useful to see how our results compare. Another modification would be a non-circular cylinder, or if it had some surface roughness. A direct solution of these problems would be challenging, however it could be solved perturbatively if the roughness was small. Moreover, the correction to the drag could be determined using the Reciprocal theorem using only the results here \cite{masoud2019reciprocal,joseph1973domain}. 

As done by Crowdy \cite{crowdy2011treadmilling}, these solutions could also be combined with the reciprocal theorem to study the hydrodynamic interactions of two-dimensional squirming swimmers. In fact, during the reviewing process of this paper we were made aware of related work that appeared unpublished in a PhD thesis on two-dimensional swimmers \cite{warwick87268}. There are  similarities between that work and ours, namely, they also found the Goursat functions for non-touching cylinders. However, no comment was made about the touching limit, and there was no mention was made of topological defects in the velocity field.

\section*{Acknowledgements}
We thank Jens Eggers and Tanniemola Liverpool for a careful reading of the manuscript, and a reviewer who helped us understand the touching limit. This work was supported by EPSRC studentship.

\setcounter{equation}{0}
\setcounter{section}{0}
\renewcommand{\theequation}{A\arabic{equation}}   
\renewcommand{\thesection}{APPENDIX \Alph{section}}  

\section{Coefficients for non-touching cylinders}
\label{sec:appen coeff}
We define the real and imaginary part of each coefficient as $A_i=\alpha_i+\ci \beta_i$. These are given by
\begin{equation*}
    \begin{gathered}
        \alpha_1=-\frac{a^2 (1 + S^2) \text{U}_1}{((-1 + a^2) (a^2 + S^2) - 2 a^2 (1 + S^2) \log{a})},\\
        \beta_1=\frac{-2 a^3 S^2 (\Omega_1 + \Omega_2) + a S^2 (S^2 \Omega_1 + \Omega_2) + a^5 (\Omega_1 + S^2 \Omega_2)}{2 (-1 + a^2) (a^2 - S^2)^2}
,\\
        \alpha_2=\frac{a^3 \text{U}_1}{(-1 + a^2) (a^2 + S^2) - 2 a^2 (1 + S^2) \log{a}}
,\\
         \beta_2=\frac{a^2 (-S^4 \Omega_1 + 2 a^2 S^2 (\Omega_1 - \Omega_2) + S^2 \Omega_2 + a^4 (-\Omega_1 + S^2 \Omega_2))}{2 (-1 + a^2) (a^2 - S^2)^2 (-1 + S^2)}
,\\
A_3= -\frac{S^2}{a^2}\overline{A}_2,\\
        \text{U}_1=\text{U}_0/2,\quad \text{U}_2=-\text{U}_0/2,\\
        \text{V}_1=\frac{(-1 + a^2) (a^2 - S^2) \beta_2}{a^3} + 2 \beta_1 \log{a},\\
        \begin{aligned}
        \text{V}_2 = &\frac{1}{a^3} (a \beta_1 - a S^2 \beta_1 + \beta_2 - 3 a^2 \beta_2 + a^4 \beta_2 + S^2 \beta_2 + a^2 S^2 \beta_2 - S^4 \beta_2 + a^2 \Omega_1 - a^2 \Omega_2 +\\
        & 2 a^3 d \Omega_2 + 2 a^3 \beta_1 \log{a} - 2 a^3 \beta_1 \log{S}).
        \end{aligned}
    \end{gathered}
\end{equation*}

\section{Coefficients for touching cylinders}
\label{sec:appen touching coeffs}
The real and imaginary parts of each coefficient $B_i=a_i+\ci b_i$ are given by
\begin{equation*}
\begin{gathered}
a_1 = \frac{(1+m)^2\Omega_1-(1+m)(1+2m)\text{V}_1}{2m^2},\\
b_1=\frac{3(1+m)(1+2m)\text{U}_1}{4m^2(3+m)},\\
a_2=\frac{3(1+m)^2\text{U}_1}{4m(3+m)},\\
b_2 = \frac{(1+m)^2(\Omega_1-\text{V}_1)}{4m^2},\\
a_3=0,\\
b_3=\frac{(1+m)^3\text{U}_1}{4m^2(3+m)}.
\end{gathered}
\end{equation*}

\bibliography{biblio}
\end{document}